\documentclass[%
 reprint,
superscriptaddress,
 amsmath,amssymb,
 aps,
pra,
]{revtex4-2}

\usepackage{graphicx}
\usepackage{dcolumn}
\usepackage{bm}
\usepackage{hyperref}
\usepackage{xurl}
\usepackage{siunitx}
\usepackage[mathlines]{lineno}
\usepackage{physics}
\usepackage{siunitx}

\usepackage{ulem} 

\usepackage[utf8]{inputenc}
\usepackage[T1]{fontenc}
\usepackage{amsmath}
\usepackage{amssymb}
\usepackage{amsthm}
\usepackage{enumitem}

\usepackage{algorithm}
\usepackage[noend]{algpseudocode} 

\algnewcommand\algorithmicinput{\textbf{Input:}}
\algnewcommand\Input{\item[\algorithmicinput]}
\algnewcommand\algorithmicoutput{\textbf{Output:}}
\algnewcommand\Output{\item[\algorithmicoutput]}
\usepackage{dsfont}
\usepackage{pdfrender}

\usepackage{afterpage}

\usepackage{braket}
\usepackage{nicefrac}

\newcommand{\opX}{\hat{X}}

\newcommand{\opZ}{\hat{Z}}

\usepackage[dvipsnames]{xcolor}

\begin{document}
\preprint{APS/123-QED}

\title{
  Constructing Arbitrary Coherent Rearrangements in Optical Lattices
}

\author{Alexander Roth}
\affiliation{Max Planck Institute of Quantum Optics, Hans-Kopfermann-Str.1, Garching D-85748, Germany}
\affiliation{Fakultät für Physik, Ludwig-Maximilians-Universität, Schellingstr. 4, München D-80799, Germany}
\author{Liyang Qiu}
\affiliation{Max Planck Institute of Quantum Optics, Hans-Kopfermann-Str.1, Garching D-85748, Germany}
\author{Timon Hilker}
\affiliation{Max Planck Institute of Quantum Optics, Hans-Kopfermann-Str.1, Garching D-85748, Germany}
\affiliation{Department of Physics and SUPA, University of Strathclyde, Glasgow G4 0NG, United Kingdom}
\author{Titus Franz}
\affiliation{Max Planck Institute of Quantum Optics, Hans-Kopfermann-Str.1, Garching D-85748, Germany}
\affiliation{Munich Center for Quantum Science and Technology (MCQST), Schellingstr. 4, München D-80799, Germany}
\author{Philipp M. Preiss}
\email{philipp.preiss@mpq.mpg.de}
\affiliation{Max Planck Institute of Quantum Optics, Hans-Kopfermann-Str.1, Garching D-85748, Germany}
\affiliation{Munich Center for Quantum Science and Technology (MCQST), Schellingstr. 4, München D-80799, Germany}

\date{\today}

\begin{abstract}
Coherent control of motional degrees of freedom of ultracold atoms in optical lattices offers a promising route towards programmable quantum dynamics with massive particles. We propose and analyze a scheme for implementing coherent rearrangement of ultracold atoms, corresponding to arbitrary unitary transformations on single-particle motional states. Exploiting an analogy between dynamics in optical superlattices and discrete linear optics, we employ the Clements scheme to systematically construct any global $N$-dimensional single-particle unitary from tunneling and phase shifts in arrays of double wells. Tunneling is controlled globally, while local operations are achieved through site-resolved potential shifts. We numerically investigate the susceptibility of the scheme to intensity noise and addressing crosstalk. We identify key subroutines enabled by this unitary construction, including the Discrete Fourier Transform and the implementation of non-native Hamiltonians. Extending the scheme to two dimensions enables all-to-all atomic rearrangement with a circuit depth that scales sublinearly with the atom number, providing a high-density and highly scalable approach to atom rearrangement.
\end{abstract}

\maketitle

\section{Introduction}

\begin{figure}[h!]
    \centering
    \includegraphics[width=\linewidth]{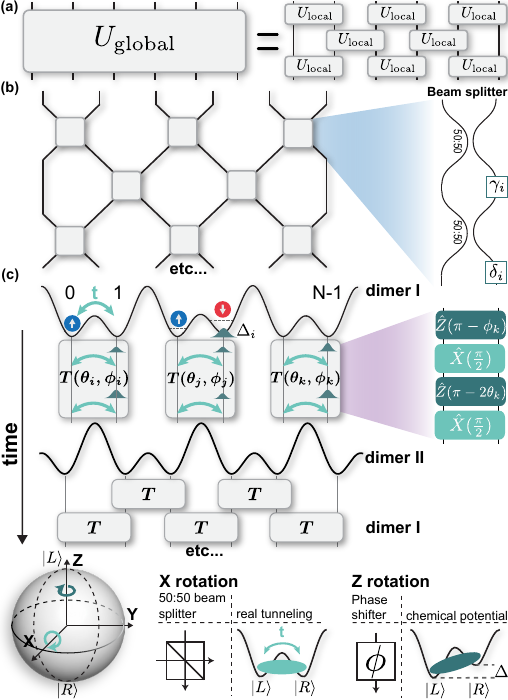}
    \caption{\textbf{Parallels of Quantum Optics and Optical Lattices. (a)} An arbitrary global unitary $U_\textrm{global}$ is systematically constructed from local two-mode unitary gates $U_\textrm{local}$. \textbf{(b)} An optical interferometer constructed using the Clements scheme realizes arbitrary $N$-mode transformations using brick-wall lattices of generalized beam splitters. The inset shows a single two-mode operation composed of two 50:50 beam splitters and phase shifters applying phases $\gamma$ and $\delta$. \textbf{(c)} Atomic dynamics between discrete modes in optical superlattices can be constructed by analogy to optical interferometers. An $\hat{X}$ rotation (beam splitter) is performed by enabling tunneling, while a $\hat{Z}$ rotation (phase shifter) is performed by applying a local energy offset $\Delta_i$. The brick-wall structure is realized by alternating the superlattice dimerization between dimer I and dimer II, which is achieved by adjusting the superlattice phase. In each time step, a two-mode transformation $T(\theta_i, \phi_i)$ is implemented in parallel on all dimers of the current dimerization through a sequence of native $\hat{X}$ and $\hat{Z}$ gates.}
    \label{fig:intro}
\end{figure}

The circuit description of quantum dynamics is a central pillar of quantum information science: Through a hierarchy of compilation levels, elementary operations can be combined into networks of gates that implement arbitrary unitary transformations~\cite{nielsen1991quantum, Barenco1995}. This abstraction separates quantum algorithms from the underlying experimental implementation and enables universal computation across a wide range of physical platforms. It also provides a systematic framework for decomposing complex dynamics into the experimentally most favorable building blocks on a given hardware architecture. For trapped ions and superconducting qubits, these building blocks correspond to finely tuned control pulses \cite{FossFeig2025, Kjaergaard2020}, while in quantum optics they are realized by discrete physical optical elements \cite{Reck1994,Clements:16, Wang2019}.

A major open challenge is to extend this compilation-based approach to the dynamics of massive mobile particles. For ultracold atomic gases trapped in optical lattices, coherent quantum motion is driven by tunneling between neighboring sites. Together with collisional interactions, such tunneling dynamics enable highly coherent evolution of thousands of particles over hundreds of tunneling times, giving rise to strongly entangled motional states \cite{Gross2017, Schaefer2020}. Despite this high degree of coherence, control over cold-atom lattice dynamics is typically limited to global operations, such as temporal modulation of lattice depths or overall confining potentials, resulting in coarse-grained control.

Bringing together the coherence of atomic motion in optical lattices with the programmability of quantum circuit models opens new perspectives for quantum simulation and computation with itinerant particles. Discrete motional unitaries can be employed for state preparation \cite{Tabares2025} and for the emulation of imaginary-time evolution \cite{Cavallar2025}. They further enable advanced measurement protocols based on collective interference \cite{Daley2012, pichler2013thermal} or on the coherent rotation of single- and two-particle correlations into the measurement basis prior to site-resolved imaging \cite{impertro2024local, schlomer2024local, mark2024efficiently}. Beyond quantum simulation, locally programmable motional unitaries form a key ingredient of recently proposed architectures for fermionic quantum computation \cite{GonzalezCuadra2023, gkritsis2025simulating} and for error-correction schemes targeting motional degrees of freedom \cite{Schuckert2024, Ott2025}.

In this article, we present a systematic framework for compiling arbitrary $N$-dimensional \textit{single-particle} unitaries from sequences of experimentally realizable local gates. This decomposition is universal. It derives from an analogy between programmable multi-mode optical beam splitters and dynamics in optical superlattices \cite{Zhang2023, impertro2024local, Zhu2025, chalopin2025optical}, where tunneling and potential shifting operations can be interpreted as beam splitter and phase-shift operations \cite{Reck1994,Clements:16}. Akin to the optical case, stroboscopic dynamics in optical superlattices can thus be programmed to transform single-particle quantum states spanning $N$ input modes into arbitrary $N$ output modes. The operation acts on single-particle motional modes in the motional ground band only and is decoupled from the particles' internal degrees of freedom. It is thus applicable to bosons and fermions in optical lattices. Experimentally, our scheme requires only globally controlled tunneling in a superlattice and local phase-shifting operations~\cite{impertro2024local}.

We highlight two key anticipated applications of the method. First, the scheme will enable the synthesis of specific unitaries as subroutines of quantum simulation approaches. We demonstrate how to engineer unitaries generated by Hamiltonians with non-native terms, such as long-range tunneling, and detail the realization of the Discrete Fourier Transform (DFT) as a gate sequence in a lattice. The DFT is a primitive for many quantum algorithms and will find important applications as a coherent mapping between position and momentum space, enabling momentum space microscopy and many-body interference experiments \cite{Daley2012, pichler2013thermal, Islam2015}. Second, we show that our approach can realize swap networks where all particles in a system perform permutations simultaneously. Applying this approach to atom rearrangement in two-dimensional $N=L\times L$ arrays, we demonstrate a scheme by which atoms in $\mathcal{O}(N)$ modes can be rearranged with all-to-all connectivity in an $\mathcal{O}(\sqrt{N})$ depth circuit. The sublinear scaling of the depth of the circuit presents a significant scaling advantage over serial, tweezer-based sorting approaches \cite{Kaufman2021}.

\section{Superlattice Architecture and Optical Analogy}
\label{chap:theory}
\subsection{Physical Platform and Motional Gates}
\label{sec:platform}

We seek to construct motional unitaries driven by tunneling dynamics in optical superlattices. Such lattices are formed by superimposing two lattices with different periodicities with dynamical control over their depths and relative spatial phases. This architecture realizes large-scale arrays of double-wells with finely tunable local potential geometries \cite{Sebbystrabley2006, chalopin2025optical, impertro2024local}. The exquisite control over dynamics within each local motif has led to powerful applications, including the realization of higher-order tunneling processes, collisional entangling gates, and local state manipulation~\cite{Anderlini2007, Foelling2007, Dai2017, Bojovic2025, Zhu2025, Zhang2023, impertro2024local}.

To establish a correspondence between optical interferometers and atoms in optical superlattices, we consider cold atoms trapped in the lowest band of a one-dimensional optical superlattice with $N$ modes as shown in Fig.~\ref{fig:intro}c. We assume a spinless gas (with the extension to several internal states being straightforward) and the non-interacting limit. The superlattice creates an array of double-well potentials with globally tunable geometry. In particular, the dimerization of the superlattice can be switched between odd and even by dynamically adjusting the relative spatial phase between the short and the long lattice \cite{Zhang2023}. Each double-well forms a natural two-level system, or motional qubit, where the atom's position in the left $\ket{L}$ or right $\ket{R}$ well defines the computational basis. Full motional qubit control is achieved via two native operations: controlling the tunneling amplitude ($t$) to perform $\hat{X}$-rotations, and applying an energy offset $\Delta$ for $\hat{Z}$-rotations. Together, these form a universal set for any local $SU(2)$ operation on a single qubit~\cite{gkritsis2025simulating, GonzalezCuadra2023}.
Both gates have already been demonstrated experimentally~\cite{Islam2015, impertro2024local}. 
The native $\hat{X}(\alpha)$ and $\hat{Z}(\theta)$ rotations are defined as:
\begin{align}
    \hat{X}(\alpha) = \begin{pmatrix}
        \cos(\frac{\alpha}{2}) & -i\sin(\frac{\alpha}{2})\\
        -i\sin(\frac{\alpha}{2}) & \cos(\frac{\alpha}{2})
    \end{pmatrix} \;;\; \hat{Z}(\theta) = \mathrm{diag}(1, e^{i\theta}).
    \label{Zasym}
\end{align}

\subsection{Unitary Decomposition via the Clements Scheme}
\label{sec:clements_scheme}
 
\begin{figure}
    \centering
    \includegraphics[width=\linewidth]{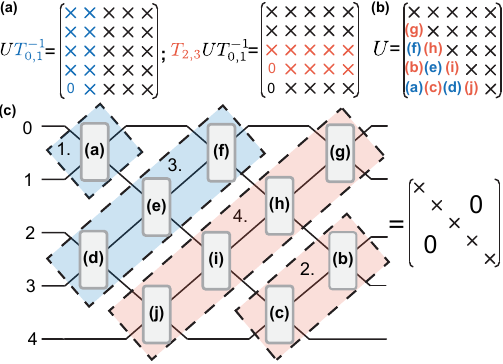}
    \caption{\textbf{Clements Scheme.} \textbf{(a)}~A Givens rotation multiplied from the right (left) mixes two columns (rows) to eliminate a specific target element of an arbitrary unitary $U$. A multiplication from the right (left) is colored in blue (red). \textbf{(b)} The Clements scheme transforms a unitary into a diagonal matrix by eliminating the sub diagonal elements in a ``snake-like'' pattern, alternating between left and right multiplied Givens rotations.\textbf{(c)}~The physical brick-wall circuit that implements the decomposition. Each rectangular gate corresponds to a single Givens rotation $T_{n,m}(\theta,\phi)$ used to nullify one element in panel (b). The different sections correspond to the subdiagonals of $U$.}
    \label{fig:Clements}
\end{figure}

To construct an arbitrary single-particle $N$-dimensional unitary transformation using only operations naturally available in a superlattice architecture, namely local $\hat{X}$ and $\hat{Z}$ operations, we leverage an analogy with linear optics, specifically the decomposition scheme developed by Clements et al.~\cite{Clements:16}. The Clements decomposition is widely used in linear optics to implement programmable multimode interferometers \cite{Bogaerts2020} that enable applications ranging from photonic quantum computing~\cite{arrazola2021quantum} and machine learning~\cite{xie2025complex} to reconfigurable microwave photonics~\cite{perez2024general}. Its key advantages are that it uses the minimal number of two-mode gates, $N(N-1)/2$, and achieves the minimal possible circuit depth of $N$.

Following the Clements scheme \cite{Clements:16}, the full unitary \mbox{$U=\sum_{i=0}^{N-1} \ketbra{\phi_i}{\psi_i}$}, which maps an initial orthonormal basis $\{\ket{\psi_i}\}$ onto a desired output basis $\{\ket{\phi_i}\}$, is constructed as a product of two-mode transformations, each acting on a two-dimensional subspace spanned by sites $n$ and $m$
\begin{align}
    T_{n,m}(\theta,\phi) = 
    \begin{pmatrix}
        \ddots &        &                      &                      &        & \\
               & 1      &                      &                      &        & \\
               &        & e^{i\phi}\cos\theta & -\sin\theta          &        & \\
               &        & e^{i\phi}\sin\theta & \cos\theta           &        & \\
               &        &                      &                      & 1      & \\
               &        &                      &                      &        & \ddots
    \end{pmatrix}_{N\times N}.
    \label{eq:Tmn}
\end{align}
and a diagonal phase matrix, $D$:
\begin{align}
    U(N) = D \left( \prod_{(n,m)\in S} T_{n,m} \right).
    \label{eq:Clementsdecomp}
\end{align}
$S$ denotes the ordered set of adjacent two-mode transformations.
This procedure is closely analogous to a QR decomposition in linear algebra, where successive orthogonal (or unitary) transformations are applied to zero out sub-diagonal elements of a matrix. In the Clements scheme, each two-mode operation $T_{n,m}$ plays the role of a localized Givens rotation that removes a specific off-diagonal element of the evolving unitary, as illustrated in Fig.\ref{fig:Clements}. Starting from the bottom-left corner, the algorithm sweeps through the sub-diagonals in a ``snake-like'' pattern: moving upward along one sub-diagonal, then downward along the next, nullifying one element at a time until the matrix becomes upper-triangular. Because $U$ is unitary, the resulting upper-triangular matrix must be diagonal. A full derivation is given in Clements et al.~\cite{Clements:16}.

In optical networks, $T_{n,m}$ represents a generalized two-mode beam splitter, which can physically be realized with two balanced beam splitters and two local phase shifters.  To implement $T_{n,m}$ in a superlattice architecture, each transformation is realized using a four-gate $Z_2X_2$ sequence (derived in Appendix \ref{sec:Z3X2proof}),  
\begin{align}
    T_{n,m}(\theta,\phi) = e^{i(\phi+\theta)} \hat{X}\left(\tfrac{\pi}{2} \right) \cdot \hat{Z}(\pi-2\theta)\cdot \hat{X}\left(\tfrac{\pi}{2}\right)\cdot \notag\\
    \hat{Z}(\pi-\phi)
    \label{eq:Z3X2}
\end{align}
The global phase accounts for the conversion from $U(2)$ to $SU(2)$ and for the phase offset between symmetric and asymmetric $\hat{Z}$ gates. The phase can, however, be absorbed by the diagonal phase matrix $D$ by finding new angles $\{\theta', \phi'\}$ (see Appendix \ref{sec: Absorption of Phase}). We will omit the prime in the following.

\begin{figure*}
    \centering
    \includegraphics[width=1.0\linewidth]{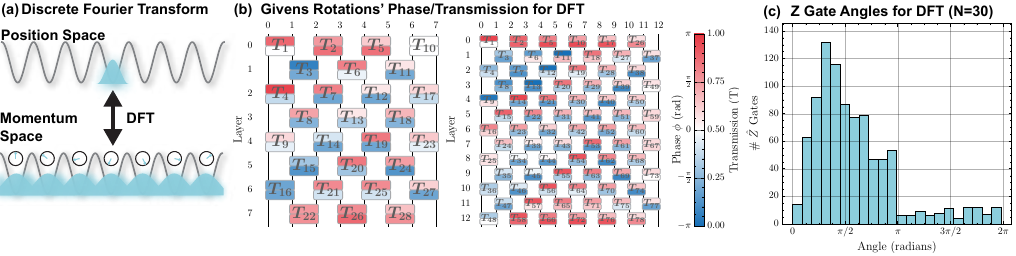}
    \caption{\textbf{Decomposition Details of a DFT.} \textbf{(a)}~A single particle, initially localized at a specific site (a position eigenstate), is coherently transformed into a delocalized state with a periodic phase structure across the entire array (a momentum eigenstate). \textbf{(b)}~The circuit decomposition of the DFT unitary for $N=8$ (left) and $N=13$ (right) systems. Each colored rectangle represents a two-mode gate. The upper rectangle indicates the gate's phase $\phi$, while the lower rectangle represents its transmission probability, $T = \cos^2(\theta)$. A translational pattern emerges for odd $N$.
    \textbf{(c)}~Histogram of the required $\hat{Z}$ gate rotation angles for an $N=30$ DFT, showing that angles are distributed across the full $[0, 2\pi]$ range.}
    \label{fig:heat maps}
\end{figure*}

As illustrated in Fig.~\ref{fig:intro}c, this sequence maps directly onto timed control pulses. Crucially, all $\hat{X}$-rotations in a given layer require the same rotation angle of $\pi/2$, meaning that all tunnel pulses can be carried out in parallel and via global control of the tunneling $t$ for a specific duration. The global $\hat{X}$ pulses are temporally interleaved with local addressing pulses applied in a deep optical lattice, which realize local chemical potential offsets and thus the local $\hat{Z}$ gates. Arbitrary local motional unitaries thus require local control over on-site phase shifts only. This architecture is inherently scalable and significantly more robust than any scheme requiring direct local control of the tunnel coupling $t$. We note that related control strategies have been proposed and demonstrated for discrete time quantum walks in spin-dependent optical lattices \cite{Robens2024}.

\section{Applications of Unitary Control}
\label{sec:use_cases}

Programmable motional unitaries of non-interacting particles in optical lattices may serve as initialization or readout tools and constitute important subroutines for information processing with itinerant particles \cite{GonzalezCuadra2023, gkritsis2025simulating, Ott2025}. Of the many possible applications, we discuss two specific instances in this section: the Discrete Fourier Transform (DFT) and the implementation of non-native Hamiltonians. 

\subsection{Discrete Fourier Transformation}

Accessing momentum-space observables is paramount for characterizing a wide range of quantum many-body phenomena, from probing band structure and spectral functions to identifying quantum phase transitions and pairing correlations~\cite{greiner2002quantum,Jotzu2014, Brown2019,  holten2022observation}.

The established procedure for probing optical lattice systems in momentum space is time-of-flight imaging, for which lattices are switched off suddenly and particles are detected in free space or in a pinning lattice after ballistic propagation, revealing momentum space densities and correlations \cite{Brown2019, holten2022observation}. The time-of-flight operation itself is incoherent, in the sense that performing the inverse operation is impractical and particles are usually not in a lowest-band lattice mode after expansion \cite{Brown2019}.

In contrast, we now consider the unitary Discrete Fourier Transform in an optical lattice. The DFT acts only within the lowest band of an optical lattice and preserves coherences between motional states. Its important features include: 1) Its fully coherent nature, allowing for coherent evolution between position and momentum space and vice-versa; 2) Its use as a subroutine for quantum algorithms, for example collective measurements on multiple system copies \cite{Daley2012, pichler2013thermal, Islam2015}; 3) The mapping of $N$ discrete lattice momentum modes to $N$ spatial modes, enabling single-shot readout of lattice momentum in quantum gas microscopes. Approximate schemes for a lattice DFT have been proposed \cite{Glaetzle2017}, but we are not aware of other exact lattice implementations.

\begin{figure*}
    \centering
    \includegraphics[width=\linewidth]{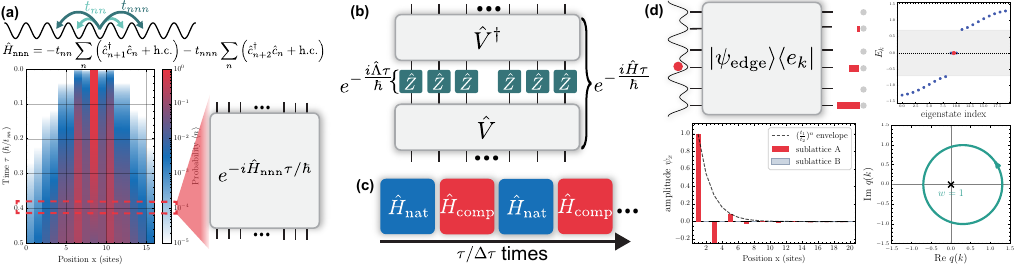}
    \caption{\textbf{Simulating Non-Native Hamiltonians and Preparing Their Eigenstates.} (a)~Direct synthesis of a non-native propagator, illustrated for a Hamiltonian with next-nearest-neighbor (NNN) hopping, with $t_{nnn}/t_{nn}=2$. The propagator up to any fixed evolution time $\tau$ (dashed box) is compiled into a single gate sequence $e^{-i\hat{H}_{nnn}\tau/\hbar}$ of depth $N$.
  (b)~Propagation via the compiled eigenbasis. The evolution time enters only through the phase layer, so propagation to arbitrary times has a fixed circuit depth.
  (c)~Non-native terms $\hat{H}_{\mathrm{comp}}$ realized by the compiled circuits are interleaved with the native lattice evolution $\hat{H}_{\mathrm{nat}}$ in a Trotter sequence of $\tau/\Delta\tau$ steps.
  (d)~Eigenstate preparation for the SSH model in the topological phase ($t_1 < t_2$). Applying the compiled eigenbasis rotation to a single localized atom $\hat{V}_{\mathrm{SSH}}\ket{e_k} = \ket{\psi_\text{edge}}$ deterministically prepares the zero-energy edge mode. The prepared amplitudes $\psi_x$ are supported on sublattice~A (red) with vanishing weight on sublattice~B and follow the expected $(t_1/t_2)^n$ envelope (dashed line). Insets: the measured energy of the prepared state lies at zero, in the center of the single-particle gap, and the trajectory of $q(k)$ encircles the origin once, corresponding to the bulk winding number $w=1$ of the topological phase.}
  \label{fig:QRW}
\end{figure*}

The matrix elements of a DFT can be written in the following way:
\begin{align}
   \mathrm{DFT} = \frac{1}{\sqrt {N}}\sum_{j,k=0}^{N-1}\omega ^{jk}\ket{k}\bra{j}
\end{align} where $\omega = e^{2\pi i / N}$ is a primitive $N$-th root of unity \cite{olson2017discrete}. In all discussions and figures, this matrix is additionally shifted, so that the zero-momentum component is centered, analogous to the \textit{fftshift}, which moves the zero-frequency component from the array edge to the center.

Figure~\ref{fig:heat maps}b shows the phase and the transmission probability ($\phi$ and $T = \cos^2(\theta)$) for the $T_{n,m}(\theta,\phi)$ of the decomposed DFTs for two system sizes. A pattern emerges for $T$ resulting from the reflection at the boundaries of the array.

The Discrete Fourier Transform implemented coherently in the lattice offers a plethora of new experimental capabilities. Its unitary nature can, for example, be used for the manipulation of specific momentum states by mapping from position space to momentum space, performing a local operation, and mapping back to position space. For fermionic many-body systems, where entropy in thermal states is concentrated in specific regions of the Brillouin zone, a single application of the DFT also localizes entropy spatially, which may enable novel cooling methods based on removing the most entropic momentum states from a system. 

Beyond the standard DFT, the decomposition framework can naturally be extended to implement the Discrete Fractional Fourier Transformation (FrFT), a unitary definition of which is provided in~\cite{candan2000discrete}. The FrFT generalizes the Fourier transform as a rotation in phase space that can prepare and measure states in a continuous set of complementary bases interpolating between position and momentum space. 

The one-dimensional DFT can be straightforwardly extended to two (or higher) dimensions by exploiting its separability. In two dimensions, this allows one to first apply a one-dimensional DFT simultaneously to all columns, followed by a one-dimensional DFT applied to all rows. Each of these one-dimensional DFTs has depth $\mathcal{O}(L)$, resulting in a total circuit depth of $\mathcal{O}(\sqrt{N})$, where we adopt the notation $N = L \times L$ with $L$ denoting the lattice size. We will retain this notation in the two-dimensional analysis, using $N$ to represent the total number of modes.

\subsection{Simulating Non-Native Hamiltonians and Preparing Their Eigenstates}

Further applications of our decomposition framework include the simulation of Hamiltonians with non-native unitary dynamics $\mathcal{\hat{U}}(\tau) = \exp\!\left(-i\hat{H}\tau/\hbar\right)$. Since the decomposition compiles any $N$-dimensional single-particle unitary, both the propagator $\mathcal{\hat{U}}(\tau)$ itself and the eigenbasis transformation of the target Hamiltonian are accessible as gate sequences. This enables a broad range of operations, of which we highlight four in Fig.~\ref{fig:QRW}.

\paragraph{Direct synthesis.}
The most immediate mode of operation is to compile the full propagator $\mathcal{\hat{U}}(\tau)$ for a fixed evolution time $\tau$ into a single gate sequence of depth $N$. We consider a Hubbard model beyond the tight-binding limit, incorporating next-nearest-neighbor (NNN) hopping with $t_{nnn}/t_{nn} = 2$ (see Fig. \ref{fig:QRW}a). NNN hopping is widely considered a necessary ingredient for stabilizing $d$-wave superconductivity~\cite{qin2020absence}.

\paragraph{Propagation via the eigenbasis.}
Following a scheme demonstrated for photonic simulators of quantum field dynamics~\cite{d2026configurable}, the time evolution can be generated by a rotation into the eigenbasis of the target Hamiltonian. For a single-particle Hamiltonian written in terms of its eigenbasis transformation $\hat{V}$, $\hat{H} = \hat{V}\hat{\Lambda}\hat{V}^\dagger$ with $\hat{\Lambda} = \mathrm{diag}(E_k)$ the single-particle eigenenergies, the time evolution allows an efficient implementation,
\begin{equation}
  e^{-i\hat{H}\tau/\hbar} \;=\; \hat{V} e^{-i\hat{\Lambda}\tau/\hbar} \hat{V}^\dagger.
  \label{eq:eigenprop}
\end{equation}
As shown in Fig.~\ref{fig:QRW}b, the propagator factorizes into three layers: the compiled rotation $\hat{V}^\dagger$ into the eigenbasis, a diagonal phase imprint $e^{-i\hat{\Lambda}\tau/\hbar}$, realized as local $\hat{Z}(-E_k\tau/\hbar)$ gates; and the compiled rotation $\hat{V}$ back. The evolution time $\tau$ enters only through the phase-imprint layer, so propagation to arbitrary times has a fixed circuit depth.

\paragraph{Interleaving with native evolution.}
To access Hamiltonians that combine non-native single-particle terms with the system's native processes, the compiled gate sequences can be interleaved with the native evolution of the lattice in a Trotter expansion, as sketched in Fig.~\ref{fig:QRW}c. Splitting the target Hamiltonian as $\hat{H} = \hat{H}_{\mathrm{nat}} + \hat{H}_{\mathrm{comp}}$, where $\hat{H}_{\mathrm{nat}}$ is generated directly by the hardware and $\hat{H}_{\mathrm{comp}}$ is the non-native part realized by our scheme, the full evolution is approximated as
\begin{equation}
  e^{-i\hat{H}\tau/\hbar} \;\approx\;
  \left( e^{-i\hat{H}_{\mathrm{nat}}\Delta\tau/\hbar}\,
         e^{-i\hat{H}_{\mathrm{comp}}\Delta\tau/\hbar} \right)^{\tau/\Delta\tau}.
  \label{eq:trotter}
\end{equation}

\begin{figure*}
    \centering
    \includegraphics[width=\linewidth]{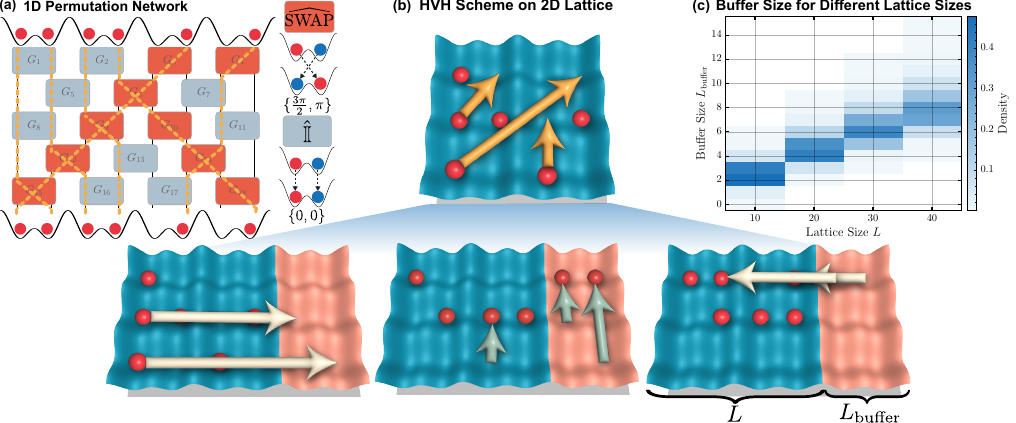}
    \caption{\textbf{Atom Rearrangement in 1D and 2D} \textbf{(a)}~A 1D permutation is decomposed into a network of two-site gates that either swap modes (red) or apply the Identity (gray). A $\widehat{\text{SWAP}}$ gate is realized by setting $\{\theta, \phi\} = \{\tfrac{3\pi}{2},\pi\}$ while the Identity $\hat{\mathbb{I}}$ is realized through $\{\theta, \phi\} = \{0,0\}$ in Eq.~\ref{eq:Z3X2}.
\textbf{(b)}~The "Horizontal-Vertical-Horizontal" (HVH) scheme is used for conflict-free 2D atom rearrangement. The three steps are: (1)~a horizontal sort moves atoms into a temporary buffer region (orange); (2)~a vertical sort places atoms in their correct target rows; and (3)~a final horizontal placement moves them into the desired target columns.
\textbf{(c)}~Distribition of the buffer size $L_{\text{buffer}}$ required for a conflict-free rearrangement of an $L \times L$ fully filled lattice, over 100,000 random targets. The colour gives the relative frequency of each outcome, normalized within each column. The worst case of $L-1$ occurs for only a vanishing fraction of configurations.}
    \label{fig:2Drear}    
\end{figure*}

\paragraph{Eigenstate preparation.}
The compiled eigenbasis rotation serves a second purpose beyond dynamics. Applied to an atom localized on a single site, $\hat{V}$ deterministically maps the position eigenstate onto the $k$-th eigenstate of the target Hamiltonian, $\hat{V}|e_k\rangle = |\psi_k\rangle$. Any single-particle eigenstate is thus prepared in a single circuit of depth $N$, without adiabatic ramps and independent of the initial state's overlap with the target. As a paradigmatic example, we consider the SSH model, $\hat{H}_{\mathrm{SSH}} = -\sum_n t_n \hat{c}^\dagger_n \hat{c}_{n+1} + \text{h.c.}$ with alternating hopping amplitudes $t_n \in \{t_1, t_2\}$. The dimerized hopping pattern is native to the superlattice architecture, and its topological edge states have been prepared via adiabatic ramps and quench protocols~\cite{meier2016observation}. Our scheme realizes these edge states at a fixed circuit depth by compiling the SSH eigenbasis $\hat{V}_{\mathrm{SSH}}$ and applying it to a single localized atom, the zero-energy edge mode is prepared exactly (Fig.~\ref{fig:QRW}d).

These examples are illustrative rather than exhaustive. Any Hamiltonian whose single-particle sector is known is accessible through either its propagator or its eigenbasis, without hardware modifications. Combined with the interactions native to the lattice, these operations broaden the class of models accessible to hybrid and fully digital quantum simulators.

\section{Atom Rearrangement}
\label{sec:atomrearrangement}

A key ingredient of the recent successes of quantum information science with neutral atoms is the ability to dynamically reconfigure atoms in optical trap arrays \cite{Kaufman2021}. This capability enables the deterministic assembly of atomic registers from stochastically loaded arrays as a starting point for quantum many-body physics or quantum computation \cite{Barredo2016, Young2024, Gyger2024, Manetsch2025}. In this context, a specific initial configuration is converted into a target configuration, although it is immaterial which atom is moved to a specific target site, granting some freedom in the design of the best routing path. A more challenging situation arises for coherent mid-circuit rearrangement, as has been used for logical gates in zoned computing architectures \cite{bluvstein2024logical}. Here, atoms have to be moved from specific initial to specific target sites while maintaining possible entanglement to static atoms in order to enable the desired computational connectivity. In all cases, rearrangement operations typically move individual or blocks of atoms in mobile tweezers and are carried out serially \cite{Kaufman2021}. Parallelized, constant-time protocols have recently been demonstrated for short-range atom rearrangement~\cite{Lin2025}. 

The scheme we outline here can be used to perform atom rearrangement using coherent ground band dynamics in optical lattices instead. As we demonstrate below, this gives rise to an appealing alternative paradigm for parallelized two-dimensional atom reconfiguration, where the total depth of the rearrangement circuit is sublinear in the number of atoms to be transported. 

The principal ingredient of the coherent rearrangement method is shown in Fig.~\ref{fig:2Drear}a. For specific phase angles, the beam splitter unitary from Eq.~\eqref{eq:Z3X2} realizes transport primitives in each double-well --- the identity $\hat{\mathbb{I}}$  for $\{\theta, \phi\}=\{0,0\}$ and the $\widehat{\text{SWAP}}$ for $\{\theta, \phi\}=\{\tfrac{3\pi}{2},\pi\}$. These circuits can be combined to form SWAP networks to rearrange atoms in one dimension, as is routinely done in quantum computing architectures with local connectivity \cite{moses2023race, wu2024modular}. Owing to the parallel implementation of $\hat{\mathbb{I}}$ and $\widehat{\text{SWAP}}$ gates in optical superlattices, a brick-wall circuit of depth $N$ can perform arbitrary, one-dimensional, parallel rearrangement of atoms in $N$ modes. 

\newcommand{\Oh}[1]{\mathcal{O}(#1)}

Extending the parallel lattice swap network to two dimensions enables the implementation of arbitrary permutation matrices on an $L \times L$ lattice. However, avoiding configurations where two atoms simultaneously occupy the same lattice site requires careful scheduling of the individual moves. To this end, we introduce the "Horizontal-Vertical-Horizontal" (HVH) scheme shown in Fig.~\ref{fig:2Drear}b which decomposes the 2D problem into three sequential 1D permutations: a conflict-resolving horizontal pass, a vertical sorting pass, and a final horizontal placement.
In the first horizontal permutation, atoms in each row are moved from their starting positions $(r_S, c_S)$ to intermediate positions $(r_S, c_I)$ chosen such that the subsequent vertical permutation is conflict-free, i.e., no two atoms in the same column are routed to the same target row.
To achieve this, the available columns are extended from the original $L$ columns to $L_{\text{ext}} = L + L_{\text{buffer}}$. In the vertical permutation, all atoms are then shifted to their respective target rows $r_T$. Finally, during the second horizontal permutation, atoms are placed in their target columns, reaching their final positions $(r_T, c_T)$.

This approach guarantees conflict-free routing for arbitrary permutations in two dimensions, provided enough buffer columns are available. The worst-case scenario requires $L-1$ buffer columns. This makes the total width $L_{\text{ext}}$ at worst $2L-1$. We explore the typical buffer size required by simulating random all-to-all permutations on an $L \times L$ array. 
Figure~\ref{fig:2Drear}c shows how often each buffer width is required for conflict-free rearrangement at different lattice sizes $L \times L$, 
demonstrating that typically required buffer sizes are much smaller than the maximum of $L-1$ auxiliary columns. The latter is needed exponentially rarely, even for a fully filled lattice. 
The plot was generated by implementing a greedy algorithm for determining the movements of the atoms.

The HVH scheme thus decomposes the rearrangement of $L^2$ modes into two sequential horizontal permutations of depth $\mathcal{O}(2L)$ and a vertical permutation of depth $\mathcal{O}(L)$, resulting in a total circuit depth of $\mathcal{O}(5L)$ for the two-dimensional reconfiguration. This corresponds to a sublinear scaling in the total system size $N=L^2$, namely $\mathcal{O}(\sqrt{N})$. Compared to the serial rearrangement typically employed in tweezer arrays, which takes at least $\mathcal{O}(N)$ moves~\cite{dudinets2025all, kaufman2015entangling, bluvstein2024logical}, the HVH approach is markedly more efficient. 

We note that for specific instances of the rearrangement problem, more efficient circuits than the HVH scheme are possible. In the most general case of $N$ distinguishable particles, $N!$ rearrangements are possible. Therefore, the number of gates required scales as $\mathcal{O}(\log_2(N!))~\approx~\mathcal{O}(N\log_2(N))$ using Stirling's approximation. The HVH scheme, on the other hand, requires $\mathcal{O}(3N\sqrt{N})$ gates.

\begin{figure*}
    \centering
    \includegraphics[width=\linewidth]{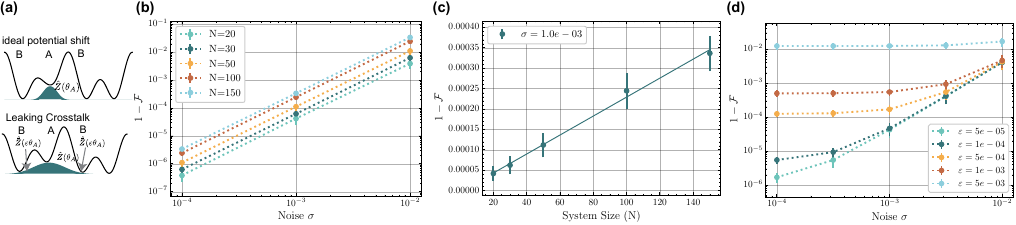}
    \caption{\textbf{Error Analysis for DFT Matrix.} Performance under two primary noise channels: random multiplicative noise on local gate angles and correlated crosstalk errors from addressing beams. The infidelity, defined as $1-\mathcal{F}$ where $\mathcal{F}~=~\lvert\langle\psi_{\text{ideal}}\vert\psi_{\text{noisy}}\rangle\rvert^2$ is the state fidelity per mode, is used as the error metric.
\textbf{(a)}~Top: Intensity fluctuations in the addressing beams result in random multiplicative phase noise on the local $\hat{Z}$ gates. Bottom: A crosstalk error occurs when an addressing beam on A affects the neighboring sites B which results in a slight tilt of the neighboring double-well. This effectively introduces a $\hat{\bar{Z}}(\varepsilon\theta_A)\equiv \textrm{diag}(e^{i\varepsilon\theta_A}, 1)$ gate on the neighboring sites B where $\varepsilon$ is the crosstalk strength.
\textbf{(b)}~Infidelity as a function of multiplicative noise strength $\sigma$ for various system sizes $N$. Each point is the average over 100 random input states and 100 noise instances. Dotted lines are guides for the eye connecting the data points.
\textbf{(c)}~Infidelity versus system size $N$ for a fixed noise strength of $\sigma=10^{-3}$. The solid line represents a linear fit to the data, demonstrating that the infidelity scales linearly with $N$.
\textbf{(d)}~Infidelity versus noise strength $\sigma$ in the presence of crosstalk error $\varepsilon$ (for $N=20$). For large crosstalk ($\varepsilon \ge 5 \times 10^{-4}$), a fidelity floor appears, limiting performance regardless of the intrinsic noise $\sigma$. Dotted lines are guides for the eye. Error bars denote the standard deviation.}
    \label{fig:errordft}
\end{figure*}

The coherent rearrangement scheme allows any atom to be moved to any site on demand, enabling all-to-all connectivity in optical lattice systems. Such dynamic reconfigurability is essential for a wide range of advanced quantum protocols, from implementing non-native couplings in quantum annealing~\cite{lechner2015quantum} to generating large-scale cluster states for measurement-based quantum computing~\cite{pham20122d}. It also represents a critical capability for fault-tolerant architectures, where qubits must be dynamically rearranged for logical gate operations and error correction~\cite{bluvstein2024logical}.

Our approach differs from rearrangement by physical transport in moving tweezers in its dominant error channel. The latter places atoms at the correct target site with high fidelity, but often introduces motional excitations to higher bands, requiring additional cooling after transport \cite{Kaufman2021, bluvstein2024logical, Gyger2024, Young2024}. In contrast, the coherent rearrangement protocol involves only lowest-band dynamics. Phase errors may cause atoms to occupy incorrect sites (see next section for a detailed error analysis), but they remain in the ground band. This error channel can be advantageous for architectures relying on collisional interactions.

Compared to tweezer-based atomic rearrangement, for which atom spacings are typically several \unit{\micro \meter} \cite{Kaufman2021, bluvstein2024logical}, the coherent rearrangement in tunnel-coupled lattices can support much higher densities of order one particle per \unit{\micro\meter\squared}. This provides interesting perspectives for neutral atom quantum computing architectures, for which it is desirable to place many atomic qubits in the field of view of a high-NA objective \cite{Manetsch2025}. High-density coherent rearrangement can significantly enhance the number of atoms that can be stored in a given register area. One can also readily imagine hybrid tweezer-lattice architectures, where mobile tweezers transport atoms over large distances between landing zones, around which atoms are rearranged block-wise in high-density lattice arrays via the coherent rearrangement presented here. Similar considerations apply to the atom-by-atom assembly of Hubbard systems in optical lattices, where one of the central technical challenges is the placement of atoms in lattices at sub-\unit{\micro \meter} scales~\cite{Young2024, Gyger2024}.

\section{Error Analysis}
\label{chap:error}

\begin{figure}
    \centering
    \includegraphics[width=\linewidth]{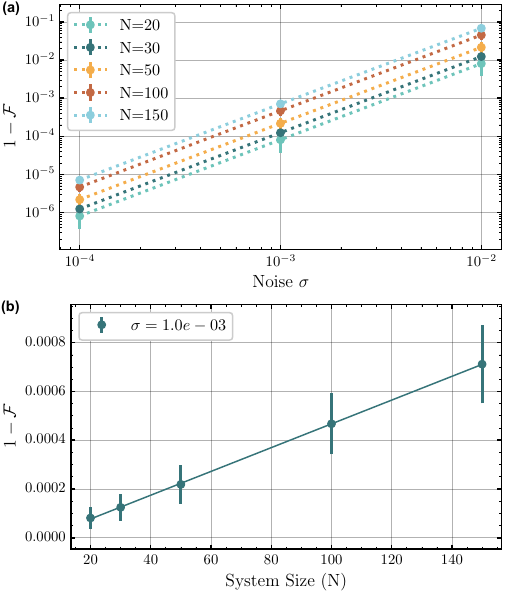}
    \caption{\textbf{Error Analysis for Atom Rearrangement.} \textbf{(a)} The fidelity was averaged over 50 random permutation matrices, each tested with 50 random initial states and 50 noise instances. \textbf{(b)} The fidelity decreases linearly with system size. A linear fit was applied.}
    \label{fig:errorperm}
\end{figure}

The realization of programmable global unitaries depends on the precision of the global tunneling $\hat{X}$ and local phase shift $\hat{Z}$ gates. Below we numerically investigate the precision required on individual gates to realize relevant applications. 

The tunneling operations $\hat{X}$ can be realized by global control of superlattice parameters, which can result in very high stabilities and accuracies of the desired gates \cite{Zhang2023, chalopin2025optical, Bojovic2025}. The local phase shifting $\hat{Z}$ is much more delicate, as it relies on high-resolution optics and precise alignment between addressing beams and optical lattices. Moreover, the dominant imperfection of the $\hat{X}$ pulse in typical experiments is a residual detuning of the double well rather than an error in the pulse area~\cite{Bojovic2025}. Such a detuning is generated by the same operator $\sum_i \Delta_i \hat{n}_i$ as the intentional $\hat{Z}$ gate, so to leading order it enters the circuit as an additional phase error and is already captured by the $\hat{Z}$ error model. We therefore assume the $\hat{X}$ gates to be error-free and ascribe all imperfections to the phase angles imprinted by the local $\hat{Z}$ gates.

We investigate three primary error channels: random multiplicative noise, correlated crosstalk affecting neighboring modes, and residual interactions.

For benchmarking the algorithm's robustness, the target is to understand the noise levels permissible to achieve a high fidelity. For concreteness, we target an infidelity $1 - \mathcal{F}<10^{-3}$, where $\mathcal{F} = |\braket{\psi_\text{ideal}|\psi_\text{noisy}}|^2$ is the state fidelity per mode.  

\subsection{Random Noise}
\label{subsec:rngnoise}
We first model imperfections in gate calibration as a multiplicative error on the target rotation angle $\theta_0$. The actually implemented angle $\theta$ is given by:
\[ \theta= \theta_0 (1 + \delta)\]
where $\delta$ is a random variable drawn from the Gaussian distribution with standard deviation $\sigma$ ($\delta~\in~\mathcal{N}(0,\sigma^2)$) for each application of the gate. This type of error corresponds to pointing fluctuations between the optical lattice and addressing optics or to power fluctuations in the addressing beams. 

Figure~\ref{fig:errordft}b shows the infidelity for implementing a DFT under this noise model. The infidelity has a power-law behavior, as seen in Fig.s~\ref{fig:errordft}b-d, and can be fitted using 
\begin{align}
    1-\mathcal{F}=CN^k\sigma^b.
    \label{eq:fit}
\end{align} 
From Fig.~\ref{fig:errordft}c, a linear behavior with $N$ can be seen, therefore, $k=1$. For the DFT, $b$ was found to be $b=1.998\pm0.010$ and $C=2.16\pm0.15$. This quadratic dependence on $\sigma$ arises because for an arbitrary angle $\theta$, the error term in the unitary $\exp(i\theta(1+\delta))$ is first-order in $\delta$, which leads to a second-order contribution to the state fidelity.

For atom rearrangement protocols we see the same quadratic noise scaling of the infidelity ($b =1.996\pm0.009$) and an error coefficient of $C = 4.3 \pm 0.3$ (see Fig.~\ref{fig:errorperm}).

\subsection{Correlated Errors}
\label{subsec:crosstalk}
A second major error source is optical crosstalk, where addressing a target site unintentionally affects its neighbors. Errors of this type occur if the intensity of the addressing beam is not fully localized to a single lattice site. 

We model this crosstalk by assuming that a $\hat{Z}(\theta_A)$ gate, intended for site~A, simultaneously induces an unwanted rotation $\hat{Z}(\varepsilon\theta_A)$ on each neighboring site~B (Fig.~\ref{fig:errordft}a). Here, $\varepsilon$ is the dimensionless crosstalk strength. Consequently, the total rotation experienced by site~B is a combination of its own noisy gate and the leakage from~A. To simulate this error, we introduce a new gate $\hat{\bar{Z}}(\theta)\equiv \textrm{diag}(e^{i\theta},1)$. This gate only acts on the site B. The resulting operation on site~B is given by:
\begin{align}
    \hat{\bar{Z}}_{\mathrm{final}}(\theta_{B}) =\hat{\bar{Z}}(\varepsilon\theta_{A_1}+\varepsilon\theta_{A_2})
\end{align}
where $\theta_{A_{1,2}}$ is the angle of the $\hat{Z}$ gate on the two neighboring sites $A_1$ and $A_2$.
The model retains only nearest-neighbour crosstalk, as the addressing intensity at the next-nearest site is much smaller than at the nearest one.

Figure~\ref{fig:errordft}d investigates the impact of this correlated crosstalk error on the fidelity by implementing a DFT for a system size of $N=20$. From the diagram, it is clear that $\varepsilon$ has to be smaller than $1\cdot 10^{-3}$ to reach a fidelity of $1-\mathcal{F}<10^{-3}$. For larger system sizes, the crosstalk error will have to decrease to even smaller values. Achieving such low crosstalk is one of the major experimental challenges of our scheme. 

\subsection{Residual Interactions}
\label{sec:residual}

The decomposition presented in this paper rests on the correspondence between superlattice dynamics and linear optics. Residual on-site interactions present during the gate pulses perturb this picture. Residual interactions enter through a single term of the Hubbard Hamiltonian, $U_\textrm{int}\hat{D}=\frac{U_\textrm{int}}{2} \sum_{i=1}^N \hat{n}_i(\hat{n}_i - 1)$ which is quartic in the mode operators and thus breaks the linear-optics analogy. It vanishes identically for a single particle and for spinless fermions, and contributes only when two or more atoms transiently share a well.

\begin{figure}
    \centering
    \includegraphics[width=\linewidth]{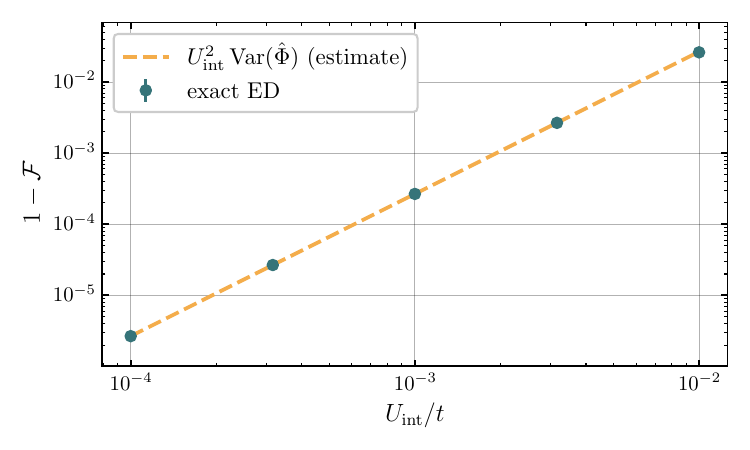}
    \caption{\textbf{Error Analysis for Residual Atom Interaction.} Infidelity of a DFT as a function of the residual on-site interaction strength $U_\textrm{int}/t$. The output state is obtained by exact evolution of the full pulse sequence including interactions. The reference is the $U_\textrm{int}=0$ evolution of the same circuit. Markers show the result for an $N=10$ site array with $n=10$ bosons with error bars marking the standard deviation. The dashed line is the first-order dephasing estimate $U_\textrm{int}^2\,\mathrm{Var}(\hat{\Phi})$.}
    \label{fig:interaction}
\end{figure}

We evaluate this channel exactly, by evolving the full pulse sequence in the fixed-particle-number Fock space with an interaction term $U_\textrm{int}\hat{D}$ included during the $\hat{X}$ and $\hat{Z}$ gates and comparing the output to the $U_\textrm{int}=0$ evolution of the same circuit. Figure~\ref{fig:interaction} shows the resulting infidelity for a DFT with $N=10$ and $n=10$. A power-law fit $1-\mathcal{F}=A(U_\textrm{int}/t)^b$ yields $b=1.992\pm0.004$ and $A=515\pm14$.

The quadratic exponent has a simple origin. To leading order in $U_\textrm{int}$ the interacting evolution factorizes as $\lvert\psi_U\rangle=\mathcal{U}_0\exp(-iU_\textrm{int}\hat{\Phi})\lvert\psi_0\rangle$, with $\mathcal{U}_0$ the ideal propagator and
\begin{align}
  \hat{\Phi} = \frac{1}{\hbar}\int_0^{\tau_\text{tot}}
               \mathcal{U}_0^\dagger(\tau')\,\hat{D}\,\mathcal{U}_0(\tau')\,d\tau'
  \label{eq:doublon}
\end{align}
the interaction-picture doublon phase accumulated along the ideal trajectory. A uniform phase is global and leaves the fidelity unchanged, so the mean $\langle\hat{\Phi}\rangle$ cancels and the dephasing is set by the variance of $\hat{\Phi}$ across the many-body trajectory,
\begin{align}
  1-\mathcal{F} \simeq U_\textrm{int}^2\,\mathrm{Var}(\hat{\Phi})
  = U_\textrm{int}^2\big(\langle\hat{\Phi}^2\rangle-\langle\hat{\Phi}\rangle^2\big)
  \propto(U_\textrm{int}/t)^2 .
  \label{eq:phi_int}
\end{align}
The variance estimate reproduces the exact points across the plotted range (dashed line in Fig.~\ref{fig:interaction}). As $\hat{\Phi}$ is defined through the ideal propagator $\mathcal{U}_0$, it is evaluated for any given problem instance directly from the non-interacting dynamics. The residual-interaction error therefore always scales as $(U_\textrm{int}/t)^2$ with only the prefactor depending on the target unitary. This allows the error to be estimated in regimes where exact evolution is numerically intractable.
 
Maintaining an infidelity $1-\mathcal{F}<10^{-3}$ for a DFT on $N=10$ modes with $n=10$ bosons therefore requires a residual interaction $U_\textrm{int}/t \lesssim 1.36\times10^{-3}$. Such values are within experimental reach by tuning the scattering length toward a zero crossing via a Feshbach resonance, or by working at reduced filling. We note that, in contrast to higher-band heating in tweezer transport, this error channel preserves the lowest-band character of the dynamics: interactions redistribute population among lattice sites but do not couple to higher bands, so the failure mode is a coherent misrouting rather than incoherent loss.

\section{Experimental Implementation}
\label{sec:experiment} 

Implementing the scheme imposes two competing requirements on the lattice. Tunneling must be fast enough for a circuit of depth $N$ to fit within the motional coherence time, while the lattice spacing must be large enough to allow site-resolved local $\hat{Z}$ gates. At fixed dimensionless lattice depth $V_0/E_R$, the tunneling amplitude is a fixed fraction of the recoil energy $E_R = h^{2}/8ma^{2}$. The duration of a global $\hat{X}(\pi/2)$ pulse therefore scales as 
\begin{equation}
    \tau_X = \frac{\pi \hbar}{4 t} \propto \frac{1}{E_R} \propto m a^{2},
    \label{eq:pulse_time}
\end{equation}
such that both a large atomic mass and a large lattice spacing reduce the gate speed. Site-resolved addressing, by contrast, favors a lattice spacing comparable to the addressing wavelength $\lambda_{\mathrm{addr}}$. A light atomic species in a wide-period lattice provides a natural compromise, which can, for example, be realized using shallow-angle superlattices, whose period is set by the intersection angle of the interfering beams rather than directly by the wavelength. Such lattices can, for example, realize spacings of order \SI{1}{\micro\meter}, which for $^{6}$Li correspond to recoil energies of several kilohertz~\cite{chalopin2025optical}.

Within such a lattice, the lattice depth serves as a dynamical knob that spans three orders of magnitude in tunneling, thereby providing a natural separation between the two native operations. At $V_0 \approx 50\,E_R$ the tunneling is effectively frozen, $t/h < 1\, \rm{Hz}$, and the phase acquired from residual tunneling during a phase pulse is of order $10^{-4}\,\mathrm{rad}$. The local $\hat{Z}$ gates are applied in this regime, where a potential shift can neither detune nor reshape a tunneling pulse, allowing the two gate types to be calibrated independently. Lowering the short lattice to a few recoil energies sets the intra-dimer tunneling to a typical value of several hundred Hz~\cite{chalopin2025optical}. 
Importantly, Eq.~\eqref{eq:pulse_time} does not impose a hard limit on the tunneling rate: it applies to evolution confined to the lowest band, whereas pulse shaping that permits transient occupation of higher bands can realize the same transfer faster~\cite{Singh2025, Reuter2026, Weil2026}. By comparison, the phase layers are considerably faster. With $\eta_i$ the phase imprinted on site $i$, their duration $\tau_Z = \hbar \max_i |\eta_i| / \Delta_{\max}$ decreases with increasing available site offset $\Delta_{\max}$, and large offsets are in principle accessible: the relevant upper bound is set not by the available addressing intensity, but by the band gap, of order $10\,E_R$ at $V_0 = 50\,E_R$, beyond which the shifted site becomes resonant with the first excited band of its neighbour. Offsets an order of magnitude below this limit, $\Delta_{\max}/h \sim 10\, \rm{kHz}$, already give $\tau_Z \approx \SI{50}{\micro\second} \ll \tau_X$, leaving the overall sequence duration dominated by tunneling. A circuit of depth $N$ then takes $N(2\tau_X + 2\tau_Z)$, about $6\,$ms for the $N=10$ Fourier transform analysed above, compared with demonstrated decoherence times of tens of milliseconds~\cite{chalopin2025optical}. Reaching larger $N$ therefore requires extending the motional coherence time rather than introducing new capabilities.

The local $\hat{Z}$ gate must be both spatially selective and phase accurate. Reaching $1-\mathcal{F}<10^{-3}$ requires a crosstalk $\varepsilon\lesssim10^{-3}$ (Sec.~\ref{subsec:crosstalk}). Evaluating the light shift on the neighboring Wannier function for diffraction-limited addressing beams, we find $\varepsilon<10^{-3}$ for site separations beyond $3\,r_\mathrm{Airy}$, with $r_\mathrm{Airy}=0.61\,\lambda_\mathrm{addr}/\mathrm{NA}$. This condition can be met by judicious choice of optical system parameters, for instance with lithium atoms in a lattice with $a=$ \SI{1}{\micro\meter}, an addressing wavelength $\lambda_\mathrm{addr}$ near the the $2S\rightarrow3P$ transition of lithium at $323\,$nm, and a common  $\mathrm{NA}=0.6$. The phases themselves must be accurate at the $10^{-3}$ level identified in Sec.~\ref{subsec:rngnoise}. Global phase errors arise from fluctuations of the superlattice phase which has been stabilized to $4.5\,$mrad shot to shot~\cite{chalopin2025optical}. Local phase errors are set by the intensity and pointing stability of the addressing beam, which must therefore be stabilized at the $10^{-3}$ level.

Beyond gate speed and precision, the main limitation to scaling is static rather than dynamic: lattice-beam curvature causes the tunneling at the edge of a large array to differ from that at its centre, and the system sizes considered here already require beam waists spanning several hundred sites. Without scaling to such large lattice beams, coherent and static inhomogeneities from the lattice envelope can be compensated algorithmically: once the tunneling amplitudes and site offsets are characterized, pulse engineering and numerical optimal control can compensate for the resulting amplitude and detuning errors through calibrated control sequences~\cite{Cummins2003, Berger2024}.
 
\section{Discussion}
\label{Conclusion}

In this work, we present a framework for decomposing arbitrary $N$-dimensional single-particle motional unitaries into native local gates for quantum simulators based on optical superlattices. By translating techniques from multiport interferometry, our method provides a practical and efficient path to programming atomic dynamics in optical lattices. We show that this framework can realize essential 1D primitives such as the Discrete Fourier Transform and all-to-all atom rearrangement. Extending these ideas to two-dimensional systems with auxiliary sites, we propose a highly favorable rearrangement protocol with $\mathcal{O}(\sqrt{N})$ circuit depth that quadratically outperforms naive serial approaches. Our error analysis establishes practical performance thresholds for experiments based on globally controlled tunneling and locally controlled potential bias.

We emphasize that the scheme derives from \textit{linear} optics and applies to non-interacting particles, or to the regime where residual interactions are weak enough to act as a collisional phase error, which we bound in
Sec.~\ref{sec:residual}. 
Stronger interactions break the analogy and require a full many-body description.

Our approach describes the unitary transformation of a set of single-particle modes that remain orthogonal throughout their evolution \cite{Reck1994, Clements:16}. These transformations are independent of the nature of the particles occupying the modes. In particular, the scheme is not affected by Pauli blocking and applies to bosons and fermions. The statistics of the underlying particles only play a role in the number correlations observed at the output, in close analogy to optical many-body interferometers and boson sampling \cite{Brod2019}.

The correspondence between lattice dynamics and interferometers may be further expanded upon by using even more compact gate fabrics~\cite{Fldzhyan2020, Bell2021} or by decomposing atomic dynamics into building blocks of larger size \cite{Arends2024, Yasir2025}. Looking ahead, our framework opens the door for the programmable many-body interference of bosonic and fermionic ultracold atoms \cite{Tichy2012, Young2024}. It further provides a direct path towards momentum-space microscopy, where unitary transformation between momentum and position space enable the manipulation and readout of individual momentum modes. The scheme also enables probes of long-range coherence and pairing correlations \cite{schlomer2024local, mark2024efficiently}, for example in fermionic gases with momentum-space pairing signatures \cite{holten2022observation}. Finally, the ability to implement tailored unitary maps allows for the simulation of non-local Hamiltonians and quantum walks on arbitrary graphs \cite{Childs2004}. 

\begin{acknowledgments}

We acknowledge helpful conversations with Andrea Alberti, Immanuel Bloch, Robert Ott, Hannes Pichler, Torsten Zache, Johannes Zeiher and the FermiQP Team. This work was supported by the Max Planck Society (MPG), the German Federal Ministry of Research, Technology and Space (BMFTR grant agreement 13N15890, FermiQP), Germany's Excellence Strategy (EXC-2111-390814868) and the European Union’s Horizon 2020 research and innovation program (grant agreements No 948240 — ERC Starting Grant UniRand -- and No 101212809 -- ERC Proof of Concept FermiChem) and the Horizon Europe program HORIZON-CL4-2022 QUANTUM-02-SGA (project 101113690, PASQuans2.1). T.H. received funding from the European Research Council (ERC) under the European Union’s Horizon Europe research and innovation program (Grant Agreement No 101165353 — ERC Starting Grant FOrbQ).

\section*{Data Availability}
The data that support the findings of this article are
openly available \cite{gitlabaccess}.

\end{acknowledgments}

\begin{appendix}

\section{Derivation of the ZXZX Angles}
\label{sec:Z3X2proof}

We construct the two-mode transformation
\begin{align}
T_{n,m}(\theta,\phi)=\begin{pmatrix} e^{i\phi}\cos\theta & -\sin\theta\\[2pt]
e^{i\phi}\sin\theta & \cos\theta\end{pmatrix}
\label{eq:appA-target}
\end{align}
from two identical tunneling pulses and two local phase gates. The native gates are the $\opX$-rotation and the local phase gate
\begin{align}
\opX(\alpha)=\begin{pmatrix}\cos\tfrac{\alpha}{2} & -i\sin\tfrac{\alpha}{2}\\[2pt]
-i\sin\tfrac{\alpha}{2} & \cos\tfrac{\alpha}{2}\end{pmatrix},
\qquad
\opZ(\eta)=\begin{pmatrix}1 & 0\\ 0 & e^{i\eta}\end{pmatrix}.
\end{align}
We seek the four-gate sequence $\opX(\tfrac{\pi}{2})\,\opZ(2\theta')\,\opX(\tfrac{\pi}{2})\,\opZ(\phi')$, i.e. two equal $\tfrac{\pi}{2}$ pulses interleaved with phase gates of angle $2\theta'$ and $\phi'$ to be determined.

We first evaluate the inner three gates.
\begin{align}
\opX(\tfrac{\pi}{2})\,\opZ(2\theta')\,\opX(\tfrac{\pi}{2})
=-i\,e^{i\theta'}\begin{pmatrix}\sin\theta' & \cos\theta'\\[2pt]
\cos\theta' & -\sin\theta'\end{pmatrix}.
\label{eq:appA-refl}
\end{align}
The two equal pulses thus produce a real reflection (up to the scalar $-i\,e^{i\theta'}$), not a rotation; the rotation character of Eq.~\eqref{eq:appA-target} must therefore be supplied by the trailing phase gate. Multiplying Eq.~\eqref{eq:appA-refl} on the right by
$\opZ(\phi')$,
\begin{align}
\opX(\tfrac{\pi}{2})\,\opZ(2\theta')\,\opX(\tfrac{\pi}{2})\,\opZ(\phi')
=-i\,e^{i\theta'}\begin{pmatrix}\sin\theta' & e^{i\phi'}\cos\theta'\\[2pt]
\cos\theta' & -e^{i\phi'}\sin\theta'\end{pmatrix}.
\label{eq:appA-full}
\end{align}
We now match Eq.~\eqref{eq:appA-full} to the target Eq.~\eqref{eq:appA-target}. Matching the (1, 1) entries gives
\begin{align}
\sin\theta'=\cos\theta\quad\Longrightarrow\quad \theta'=\tfrac{\pi}{2}-\theta,
\label{eq:appA-thetap}
\end{align}
Substituting $\theta'=\tfrac{\pi}{2}-\theta$, Eq.~\eqref{eq:appA-full} becomes
\begin{align}
-i\,e^{i\theta'}\begin{pmatrix}\cos\theta & e^{i\phi'}\sin\theta\\[2pt]
\sin\theta & -e^{i\phi'}\cos\theta\end{pmatrix}.
\end{align}
For this to equal Eq.~\eqref{eq:appA-target}, the ratio of corresponding entries must be a single
common phase. From the first column this ratio is $e^{i\phi}/(-i\,e^{i\theta'})$, while from the
second column it is $-1/(-i\,e^{i\theta'}e^{i\phi'})$. Equating the two gives
\begin{align}
e^{i(\phi+\phi')}=-1\quad\Longrightarrow\quad \phi'=\pi-\phi.
\label{eq:appA-phip}
\end{align}
Collecting the results yields
\begin{align}
T_{n,m}(\theta,\phi)=e^{i(\phi+\theta)}\,
\opX(\tfrac{\pi}{2})\,\opZ(\pi-2\theta)\,\opX(\tfrac{\pi}{2})\,\opZ(\pi-\phi).
\end{align}

\section{Absorption of Phase in Diagonal Matrix}
\label{sec: Absorption of Phase}
The phase factors of the form $e^{i(\phi+\theta)}$ occurring in the $Z_2X_2$ decomposition can be moved through the gate sequence by permutation and be absorbed in an overall diagonal phase shift $D$. To see this, consider a unitary matrix $U$ decomposed into $T_{n,m}$ building blocks, e.g. for $\textrm{dim}(U)~=~4$:
\begin{align}
U= D \cdot T_{1,2}T_{2,3}T_{0,1}T_{1,2}T_{2,3}T_{0,1}.
\end{align} 
Each of these $T_{n,m}$ can be decomposed into $T_{n,m}= S*P$ where $S\in SU(2)$ and $P=\textrm{diag}(e^{i\alpha}, e^{i\beta})$. Therefore, we have 
\begin{align}
U= D\cdot (S_{1,2}P_{1,2})(S_{2,3}P_{2,3})(S_{0,1}P_{0,1})(S_{1,2}P_{1,2})\notag\\
(S_{2,3}P_{2,3}) (S_{0,1}P_{0,1}).
\label{eq:SP}
\end{align}  
For each $S_{n,m}P_{n,m}$ we can find $S_{n,m}'$ and $P_{n,m}'$ such that $S_{n,m}P_{n,m}= P'_{n,m}S_{n,m}'$.
Therefore, we can rewrite Eq.~\ref{eq:SP} as 
\begin{align}
    U= D\cdot (S_{1,2}P_{1,2})(S_{2,3}P_{2,3})(S_{0,1}P_{0,1})(S_{1,2}P_{1,2}P'_{0,1}) \notag \\(S_{2,3}P_{2,3}) (S'_{0,1})
\end{align}
since $P'_{0,1}$ commutes with $S_{2,3}P_{2,3}$. We can now combine $P_{1,2}P'_{0,1}$ into $P_{0,1,2}$. Now we repeat this and find $P'_{0,1,2}S'_{1,2}$ etc. until we end up at 
\begin{align}
    U = D\cdot P_{0,1,2,3}S'_{1,2}S'_{2,3}\dots S'_{0,1} = D'\cdot S'_{1,2}S'_{2,3}\dots S'_{0,1}
\end{align}

\section{Residual Interactions}
\label{app:interactions}
 
Throughout the sequence the atoms evolve under $\hat{H}(\tau)=\hat{H}_0(\tau)+U\hat{D}$, with $\hat{H}_0$ the ideal generator of the compiled circuit and $\hat{D}=\frac{1}{2}\sum_i\hat{n}_i(\hat{n}_i-1)$. $\hat{H}_0$ alternates between the local offsets $\sum_i\Delta_i\hat{n}_i$ of a phase-imprint stage and the hopping term $\hat{K}_p$ of a global $\hat{X}(\pi/2)$ pulse on dimer parity $p$.
 
\paragraph{Exact evaluation.} The target matrix is decomposed fully into the $Z_2X_2$ blocks of Eq.~(4)
and propagated layer by layer in the particle-number-conserving Fock space of $n$ bosons on $N$ sites. In a phase-imprint stage, site $i$ is held at a local offset $\Delta_i$ and thereby acquires the phase $\eta_i = -\Delta_i \tau_Z/\hbar$. All sites of a layer are addressed in parallel and therefore share a single duration, which we fix by assigning the largest required phase the maximum offset $\Delta_{\max}$ available to the addressing beams $|\Delta_i| = \frac{\hbar\, |\eta_i|}{\tau_Z} \le \Delta_{\max}$. Both $\sum_i\Delta_i\hat{n}_i$ and $\hat{D}$ are diagonal in the occupation basis, so they commute and the propagator is evaluated in closed form. In a tunneling stage $[\hat{K}_p,\hat{D}]\neq0$, and $\exp[-\tfrac{i}{\hbar}(\hat{K}_p+U\hat{D})\tau_X]$ is applied directly to the state vector without Trotter splitting, using \texttt{scipy.sparse.linalg.expm\_multiply}. The layer-by-layer propagation yields $\ket{\psi_U}$.
 
\paragraph{Perturbative estimate.} Writing
$|\psi(\tau)\rangle=\mathcal{U}_0(\tau)|\psi_I(\tau)\rangle$ with $\mathcal{U}_0$ the ideal propagator gives $i\hbar\,\partial_\tau|\psi_I\rangle =U\,\mathcal{U}_0^{\dagger}\hat{D}\,\mathcal{U}_0|\psi_I\rangle$. The first Magnus term then yields
\begin{align}
  |\psi_U\rangle = \mathcal{U}_0(\tau_{\rm tot})\,e^{-iU\hat{\Phi}}|\psi_0\rangle + \mathcal{O}(U^2).
  \label{eq:app_phi}
\end{align}
Since $\mathcal{U}_0$ is common to both states it cancels from the overlap, leaving $\mathcal{F}=|\langle e^{-iU\hat{\Phi}}\rangle|^2$ with $\langle\,\cdot\,\rangle\equiv\langle\psi_0|\cdot|\psi_0\rangle$. Expanding, $\langle e^{-iU\hat{\Phi}}\rangle =(1-\tfrac{1}{2}U^2\langle\hat{\Phi}^2\rangle)-iU\langle\hat{\Phi}\rangle+\mathcal{O}(U^3)$, so that $\mathcal{F}=(1-\tfrac{1}{2}U^2\langle\hat{\Phi}^2\rangle)^2+U^2\langle\hat{\Phi}\rangle^2 +\mathcal{O}(U^4)$ and
\begin{equation}
  1-\mathcal{F} \simeq U^2\,\mathrm{Var}(\hat{\Phi}).
  \label{eq:app_var}
\end{equation}
The mean enters only through the imaginary part and therefore only as a phase, which is the formal statement that a uniform collisional phase is global and leaves the fidelity unchanged; the error is set by the spread of $\hat{\Phi}$ across the many-body trajectory.
\end{appendix}
\bibliography{references}

\end{document}